# Ultrafast dynamics and fragmentation of $C_{60}$ in intense laser pulses

Zheng-Zhe Lin[1)][†] and Xi Chen[2)]

1) *School of Science, Xidian University, Xi'an 710071, P. R. China*

2) *Department of Applied Physics, School of Science, Xi'an Jiaotong University, Xi'an 710049, P.R. China*

[†] Corresponding Author. E-mail address: linzhengzhe@hotmail.com

**Abstract** - The radiation-induced fragmentation of the $C_{60}$ fullerene was investigated by the tight-binding electron-ion dynamics simulations. In intense laser field, the breathing vibrational mode is much more strongly excited than the pentagonal-pinch mode. The fragmentation effect was found more remarkable at long wavelength $\lambda \geq 800$ nm rather than the resonant wavelengths due to the internal laser-induced dipole force, and the production ratio of C and $C_2$ rapidly grows with increasing wavelength. By such fragmentation law, C atoms, $C_2$ dimers or large $C_n$ fragments could be selectively obtained by changing the laser wavelength. And the fragmentation of $C_{60}$ by two laser pulses like the multi-step atomic photoionization was investigated.

## I. INTRODUCTION

Interactions of intense ultrashort laser pulses with molecules and the competition between ionization and fragmentation have attracted considerable attentions [1-3]. Because of the highly symmetric structure, $C_{60}$ fullerene is regarded as a particular model for studying the mechanisms of molecular energy deposition and migration in intense laser field. Various experimental and theoretical studies have been carried out to explore the dynamics of $C_{60}$ in intense laser pulses [4-8], and a great deal of phenomena about the interactions between $C_{60}$ and intense laser pulses have been discovered. For example, a massive change in ionization patterns [9, 10], above-threshold ionization (ATI) [9], population of Rydberg states [11, 12], excitation of giant breathing motion [13], thermionic emission [14, 15] and high-order harmonic generation (HHG) [16-18].

Several studies have focused on the specific ions and fragments of $C_{60}$ generated in intense laser. The first systematic study was done by O'Brien *et al.* using nanosecond laser pulses. By the fragment detection of time-of-flight spectrometers [19], the primary channel of photodissociation was found to be the loss of neutral $C_2$ or $C_3$ units [20]. Then, the dynamic evolution of $C_{60}$ was observed under different laser wavelengths. By wavelengths below 1000 nm, $C_{60}$ cluster mainly gains photon energy by excited electrons out of occupied states [21]. And the dipole force plays a role in the situation of longer wavelengths [22]. For short wavelengths, the role of intermediate states in the initial process of energy deposition in large molecules has been addressed [23]. Several experimental and theoretical studies indicated that the LUMO+1 state, which can be excited through the first dipole-allowed HOMO→LUMO+1 transition and followed by coupling to electronic and vibrational degrees of freedom, plays a crucial role as doorway state in the excitation mechanism [1]. Recently, ionization and fragmentation of $C_{60}$ fullerenes via the excitation of LUMO+1 state was studied in elliptically polarized intense femtosecond laser field [24, 25] to weaken ATI and HHG by reduced electronic recollision. And molecular dynamics (MD) simulations [26, 27] were employed to generalize the rules of laser-induced fragmentation of $C_{60}$ fullerenes. However, for studying such ultrafast electronic excitation progress, MD simulation is not an appropriate theoretical approach because the motion of $C_{60}$ is beyond the Born-Oppenheimer approximation.

In this work, the interactions between $C_{60}$ fullerenes and intense laser pulses were investigated by tight-binding electron-ion dynamics (TBED) [28-31] in a wavelength range of 300~1500 nm. The breathing mode was found much more strongly excited rather than the pentagonal-pinch mode in intense radiation field. At short loser wavelength the open-cage distortion is induced by the radial stretch of the breathing mode, while the internal laser-induced dipole force plays an important role in producing C atoms or $C_2$ dimers at a wavelength longer than 800 nm. The production ratio of C and $C_2$ rapidly grows with increasing laser wavelength and the fragmentation law could be used to obtain C atoms, $C_2$ dimers or large $C_n$ fragments. In order to enhance the fragmentation efficiency, the excitation of continuous electronic transitions by two laser pulses was investigated.

## II. TBED SIMULATIONS

To investigate the interactions between $C_{60}$ and intense laser, TBED is introduced in our simulations. The electronic states in $C_{60}$ are described by the linear combination $|\psi_j> = \sum_i c_{ji} |\varphi_i>$ of covalent s and p orbitals $|\varphi_i>$ of C atoms, i.e. the column vector form

$$\psi_j = \begin{pmatrix} c_{j1} \\ c_{j2} \\ ... \\ c_{jN} \end{pmatrix}, \quad (1)$$

and the molecular Hamiltonian $\hat{H}_0$ is presented in the matrix form $H_{0ij}$, which can be derived by the non-self-consistent approach of a density-functional-based theory [32, 33]. The laser field is treated classically by vector potential $\vec{A}$ in the Coulomb gauge and the total Hamiltonian reads

$$\hat{H} = \hat{H}_0 - \frac{i\hbar e \vec{A} \cdot \nabla}{mc} + \frac{e^2 \vec{A}^2}{2mc^2}. \quad (2)$$

For laser pulse with wavelength much longer than the size of $C_{60}$, the field $\vec{A}$ is treated as uniform and the Hamiltonian matrix should be

$$H_{ij} = H_{0ij} + \frac{i\hbar e}{mc} \vec{A} \cdot \frac{\partial S_{ij}}{\partial \vec{r}_j} + \frac{e^2 \vec{A}^2}{2mc^2} S_{ij}, \quad (3)$$

where the overlap matrix element $S_{ij} = \int \varphi_i \varphi_j d^3 \vec{r}$ and $\vec{r}_j$ the position of atom carrying $\varphi_j$. For uniform field $\vec{A}$, this calculation method is rigorous rather than by the approximate Peierls substitution [28-31]. The total energy is expressed as the sum of electronic energies and a short-range repulsive pair potential [32-34]

$$E_{tot} = \sum_i (\psi_i^+ H \psi_i)/(\psi_i^+ S \psi_i) + U_{rep}, \quad (4)$$

where $H$ and $S$ the matrix of $H_{ij}$ and $S_{ij}$, respectively. Here, $H_{ij}$, $S_{ij}$ and $U_{rep}$

were calculated by Sankey and Niklewski's technique [34]. The Hellmann-Feynman force on the k-th atom reads

$$M_k \ddot{\vec{r}}_k = -\sum_i \psi_i^+ (\frac{\partial H}{\partial \vec{r}_k} - \frac{\partial S}{\partial \vec{r}_k} S^{-1} H) \psi_i / (\psi_i^+ S \psi_i) - \frac{\partial U_{rep}}{\partial \vec{r}_k}. \tag{5}$$

By a time step $\Delta t$, a unitary evolution of Schrödinger equation for $\psi_j$ is carried out by Cayley algorithm [33]

$$\psi_i(t + \Delta t) = (1 + iS^{-1} H \Delta t / 2\hbar)^{-1} \cdot (1 - iS^{-1} H \Delta t / 2\hbar) \psi_i(t), \tag{6}$$

and Eq. (5) is solved by the velocity Verlet algorithm.

Before investigating the dynamics of $C_{60}$ in intense laser field, the geometry is optimized by a damped trajectory method [35] with the electronic occupations given by the Fermi-Dirac distribution at 0 K. By a time step of $\Delta t$ =0.01 fs, the $C_{60}$ is set in an equilibrium of 300 K for 1000 fs by Riley's thermal bath [36], and then subjected to a Gaussian laser pulse

$$\vec{A} = \vec{A}_0 \sin(\omega t) \exp\left(-16(\frac{t}{T} - \frac{1}{2})^2\right) \tag{7}$$

with the duration $T$=50 fs. With the peak intensity $I_0 = \omega^2 A_0^2 / 4\pi$ and the wavelength $\lambda$ in a range of $1 \times 10^{13}$~$2 \times 10^{14}$ W/cm$^2$ and 300~1500 nm, respectively, simulations were performed 10 times for every intensity and wavelength to obtain average results.

### III. RESULTS AND DISCUSSION

The calculated bond lengths and the HOMO-LUMO gap in $C_{60}$ are 1.39 Å, 1.43 Å and 1.43 eV, which were found close to the density-functional calculation (1.41 Å, 1.46 Å and 1.67 eV) via the Perdew-Burke-Ernzerhof functional [37]. By the electronic level and oscillator strength calculations, the main peaks in the absorption spectra of $C_{60}$ locate at 638, 646 and 380 nm with relative strength 2.4:1.0:1, corresponding to the HOMO-1→LUMO, HOMO→LUMO+1 and HOMO→LUMO+2 transitions, respectively, while the HOMO→LUMO transition was found to be dipole-forbidden. To test the calculation program, simulations for the peak intensity $I_0$=1×10$^{10}$ W/cm$^2$ (much lower than the intensity for fragmentation) were performed at every wavelength $\lambda$, and the average energy of $C_{60}$ gained from the radiation field [Fig. 1(a)] shows a strong peak at 640 nm and a weak peak at 400 nm, which is in good agreement with the absorption spectra calculation. Then, the excitation of optically-active vibrational modes was investigated by the simulations for $I_0=1\times10^{12}$~$5\times10^{12}$ W/cm$^2$, in which the $C_{60}$ still remains intact. The vibrational spectrum was obtained by the Fourier transform of the velocity autocorrelation function over an interval of 1 ps following the completion of the laser pulse. For laser wavelength below 800 nm, the most noticeable feature is the excitation of the breathing mode at 390 cm$^{-1}$, especially for the absorption peak 640 nm [Fig. 1(b)]. For longer wavelength, the excitation of the pentagonal-pinch mode at 1440 cm$^{-1}$ is more remarkable than the breathing mode at low laser intensity $I_0=1\times10^{12}$ W/cm$^2$, while the latter is still dominant at $I_0=5\times10^{12}$ W/cm$^2$ [Fig. 1(c)]. The decrease in relative amplitude of the pentagonal-pinch mode with increasing laser fluence agrees with the measurements of Fleischer *et al.* [38] and the time-dependent density-functional-theory calculations of Torralva *et al.* [39].

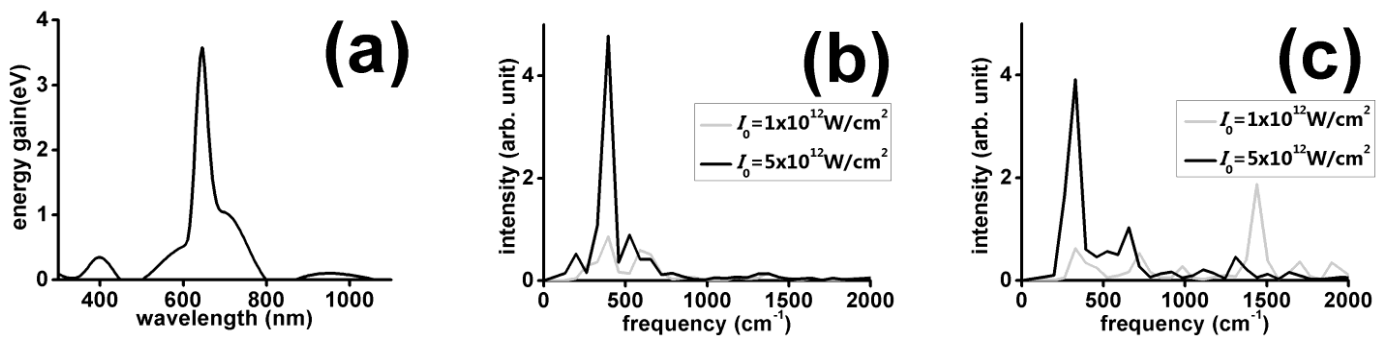

Fig. 1 The energy of $C_{60}$ gained from the radiation field by $I_0=1\times10^{10}$ W/cm$^2$ (a), and the vibrational spectrum of $C_{60}$ after being subjected to $I_0=1\times10^{12}$~$5\times10^{12}$ W/cm$^2$ laser pulses at wavelengths of 640 (b) and 1500 nm (c).

The open-cage distortion [Fig. 2(a)] or fragmentation [Fig. 2(b) and (c)] takes place when the $C_{60}$ is subjected to laser pulses of $I_0>3\times10^{13}$ W/cm$^2$. The breathing mode is remarkably excited in the radiation field, leading the rapid inflation of $C_{60}$ with increasing molecular temperature to 2000~3500 K following the laser pulse. If enough energy is gained from the radiation field, the $C_{60}$ breaks into small fragments during the violent radial stretch. In the range of $\lambda$=300~500 nm, only open-cage distortion occurs even when $I_0$ is up to $1\times10^{14}$ W/cm$^2$ [Fig. 2(a)]. For $\lambda$ near the absorption peak 640 nm, more energy is gained and the fragmentation takes place when $I_0 \geq 8\times10^{13}$ W/cm$^2$ [Fig. 2(b)], mainly producing C$_2$ dimers along with a few C atoms or small C$_n$ clusters. For longer wavelength $\lambda \geq 800$ nm, which is far away from the absorption peak, the fragmentation can even happen in lower $I_0$ with fewer C$_n$ clusters [Fig. 2(c)]. At $\lambda$=800~1064 nm, the fragmentation threshold for $I_0$ decreases from $4\times10^{13}$ W/cm$^2$ to $2\times10^{13}$ W/cm$^2$. In intense infrared radiation, the energy of C$_{60}$ gained at the resonant $\lambda$=400 or 640 nm is not remarkable due to the rapid change of electronic levels in the inflation progress. At long wavelength, the C$_{60}$ is strongly pulled by the laser-induced dipole force [40] along the electric field of linearly polarized laser and obtains more energy than that at the resonant $\lambda$. For laser pulses with lower intensity in which the C$_{60}$ keeps intact, the molecular vibration along the dipole force has been found in the quantum wavepacket simulations [5]. For higher laser intensity the C$_{60}$ breaks under the effect of the dipole force. Under the same laser intensity $I_0$, more and more C atoms rather than C$_2$ dimers are produced with increasing $\lambda$, and C$_n$ cluster can be hardly generated at long $\lambda$. For example, for $I_0$=$8\times10^{13}$ W/cm$^2$ the production of C$_2$ dimers is much higher than C atoms or C$_n$ clusters at $\lambda$=640 nm (the upper panel of Fig. 2(d)), while at $\lambda$=1064 nm the ratio of C and C$_2$ becomes 1:1.3 (the middle panel of Fig. 2(d)). At $\lambda$=1500 nm, the fragmentation produces much more C atoms than C$_2$ dimers without any C$_n$ clusters (the lower panel of Fig. 2(d)). In the electric field direction of laser, the concentration of fragments is slightly higher than other directions due to the strong laser-induced dipole force. In general, the motion of C$_{60}$ in intense laser field has an obvious relation with the wavelength. The open-cage distortion takes place in the case of short wavelength, and the fragmentation into C$_2$ dimers or C atoms takes place in long wavelength. The above result is slightly different with the simulation which deals the effect of laser as a sudden heating and obtains a continuous distribution of C$_n$ fragments [26, 27]. Such difference was also mentioned in Ref. [39]. Actually, the TBED simulation gives a more real physical picture than the simulation method of injecting energy into the C$_{60}$ molecule.

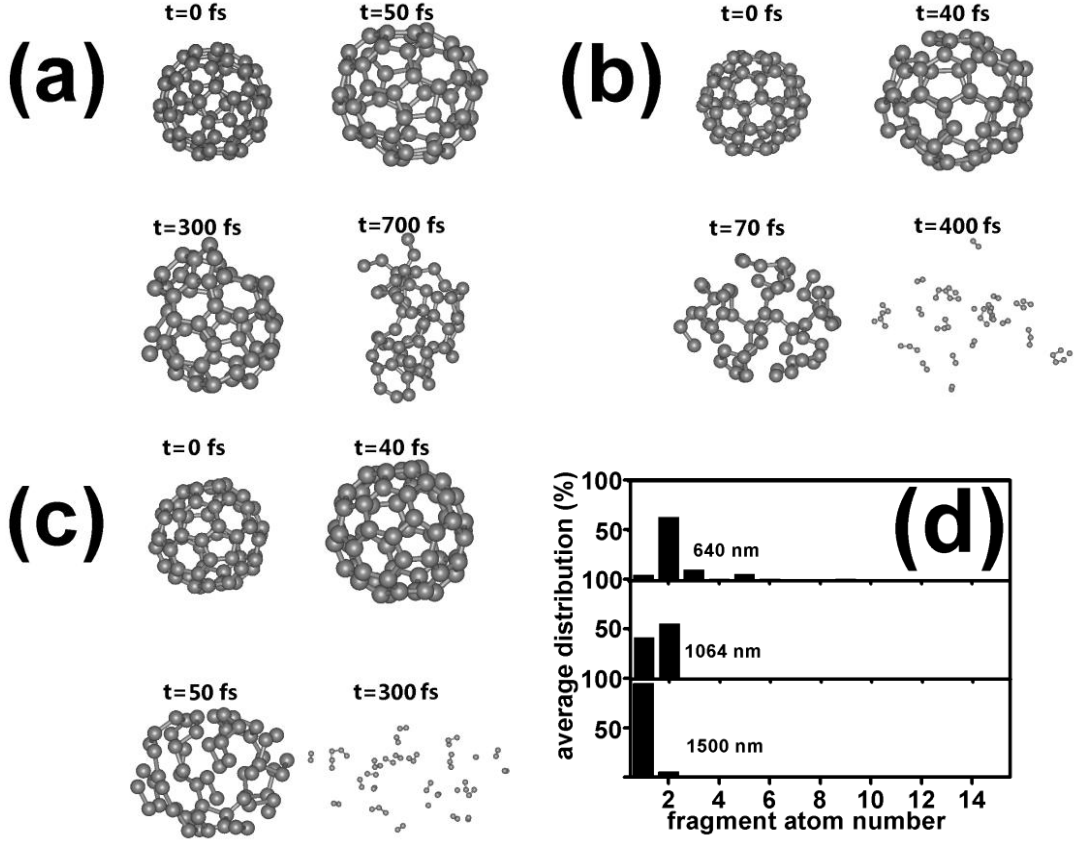

Fig. 2 Fragmentation progresses of $C_{60}$ by a laser pulse of $\lambda$=450 nm, $I_0$=1×10$^{14}$ W/cm$^2$ (a), $\lambda$=640 nm, $I_0$=8×10$^{13}$ W/cm$^2$ (b) and $\lambda$=1064 nm, $I_0$=5×10$^{13}$ W/cm$^2$ (c). For $I_0$=8×10$^{13}$ W/cm$^2$, the fragment size distribution at $\lambda$=640, 1064 and 1500 nm is shown in (d).

In order to try to find a more efficient way, $C_{60}$ fragmentation by two laser pulses was preliminarily investigated. The basic idea is to pump electrons to excited states via a resonant $\lambda_1$=640 nm and then break the $C_{60}$ by the laser-induced dipole force via $\lambda_2 \geq$800 nm, like the multi-step photoionization for atoms. The simulation system was setup by simultaneously irradiating the two laser pulses $\lambda_1$ and $\lambda_2$ with $T$=50 fs and zero phase difference on the $C_{60}$. In the range of $\lambda_2$=800~1500 nm, simulations were performed several times for every wavelength, however, no any enhancement was found. For example, by the irradiation of the first laser pulse $\lambda_1$=640 nm, $I_{0,1}$=1.5×10$^{13}$ W/cm$^2$ and the second one $\lambda_2$=1064 nm, $I_{0,2}$=1.5×10$^{13}$ W/cm$^2$, the $C_{60}$ breaks into large $C_n$ fragments with a few $C_2$ dimers [Fig. 3(a)], while more $C_2$ and small fragments are produced by one laser pulse with $\lambda$=1064 nm, $I_0$= $I_{0,1}$+ $I_{0,2}$=3×10$^{13}$ W/cm$^2$, whose intensity is the sum of the above two pulses. This may be because the $C_{60}$ is away from the resonance with $\lambda_1$ in the inflation progress due to the strong excitation of the breathing motion, and so, $\lambda_1$ does not work for the pumping. Further study could focus on the phase difference, pulse duration and time order of the two lasers.

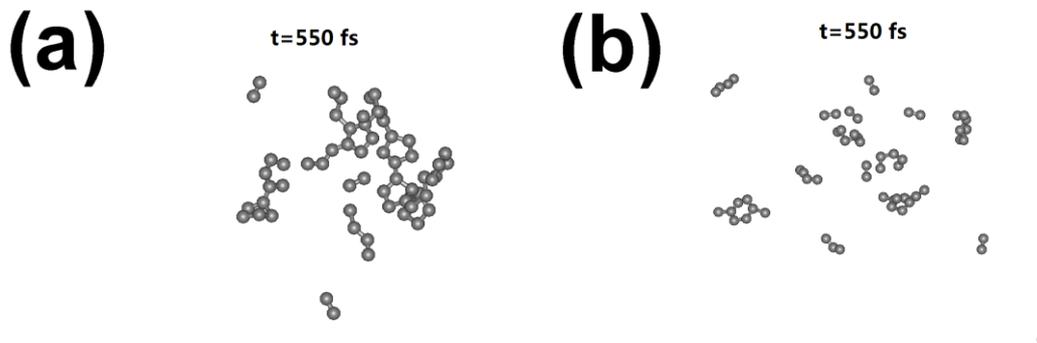

Fig. 3 Fragmentation of $C_{60}$ by two laser pulses of $\lambda_1$=640 nm, $I_0$=1.5×10$^{13}$ W/cm$^2$ and $\lambda_2$=1064 nm, $I_0$=1.5×10$^{13}$ W/cm$^2$ (a) and by a laser pulse of $\lambda$=1064 nm, $I_0$=3×10$^{13}$ W/cm$^2$ (b).

## IV. SUMMARY

In this work, TBED simulations were performed to study the fragmentation of $C_{60}$ in intense laser. The breathing mode was found much more strongly excited than the pentagonal-pinch mode in intense radiation field. Below the laser intensity for fragmentation, strong energy absorption at the wavelengths coupled with HOMO-1→LUMO, HOMO→LUMO+1 and HOMO→LUMO+2 transitions. For intense laser pulses, the fragmentation effect is more remarkable at long wavelength $\lambda \geq 800$ nm rather than the resonant wavelengths. For long wavelength, the internal laser-induced dipole force plays an important role in producing C and $C_2$ fragments, and the production ratio of C and $C_2$ rapidly grows with increasing laser wavelength. Such fragmentation law could be used to obtain C atoms, $C_2$ dimers or large $C_n$ fragments by changing the laser wavelength. By simultaneously irradiating the two laser with a same pulse duration and zero phase difference, the enhancement of fragmentation efficiency by the multi-step excitation was not found since the $C_{60}$ is away from the resonance with the laser pulse due to its strong motion in intense laser field.

**Acknowledgements**

This work was supported by the Fundamental Research Funds for the Central Universities.

**References:**

[1] A. N. Markevitch, D. A. Romanov, S. M. Smith, H. B. Schlegel, M. Y. Ivanov, and R. J. Levis, *Phys. Rev. A* **69**, 013401 (2004).

[2] S. I. Chu and D. A. Telnov, *Phys. Rep.* **390**, 1 (2004).

[3] A. N. Markevitch, D. A. Romanov, S. M. Smith, and R. J. Levis, *Phys. Rev. A* **75**, 053402 (2007).

[4] I. V. Hertel, T. Laarmann, and C. P. Schulz, *Adv. At. Mol., opt. Phys.* **50**, 219 (2005).

[5] R. Sahnoun, K. Nakai, Y. Sato, H. Kono, Y. Fujimura, and M. Tanaka, *Chem. Phys. Lett.* **430**, 167 (2006).

[6] K. K. Nakai, H., Y. Sato, N. Niitsu, R. Sahnoun, M. Tanaka, and Y. Fujimura, *Chem. Phys.* **338**, 127 (2007).

[7] M. Ruggenthaler, S. V. Popruzhenko, and D. Bauer, *Phys. Rev. A* **78**, 033413 (2008).

[8] G. P. Zhang and T. F. George, *Phys. Rev. B* **76**, 085410 (2007).

[9] E. E. B. Campbell, K. Hansen, K. Hoffmann, G. Korn, M. Tchaplyguine, M. Wittmann, and I. V. Hertel, *Phys. Rev. Lett.* **84**, 2128 (2000).

[10] M. Tchaplyguine, K. Hoffmann, O. Dühr, H. Hohmann, G. Korn, H. Rottke, M. Wittmann, I. V. Hertel, and E. E. B. Campbell, *J. Chem. Phys.* **112**, 2781 (2000).

[11] M. Boyle, K. Hoffmann, C. P. Schulz, I. V. Hertel, R. D. Levine, and E. E. B. Campbell, *Phys. Rev. Lett.* **87**, 273401 (2001).

[12] M. Boyle, T. Laarmann, K. Hoffmann, M. Heden, E. E. B. Campbell, C. P. Schulz, and I. V. Hertel, *Eur. Phys. J. D* **36**, 339 (2005).

[13] T. Laarmann, I. Shchatsinin, A. Stalmashonak, M. Boyle, N. Zhavoronkov, J. Handt, R. Schmidt, C. P. Schulz, and I. V. Hertel, *Phys. Rev. Lett.* **98**, 058302 (2007).

[14] F. Lepine and C. Bordas, *Phys. Rev. A* **69**, 053201 (2004).

[15] E. E. B. Campbell, K. Hoffmann, and I. V. Hertel, *Eur. Phys. J. D* **16**, 345 (2001).

[16] M. F. Ciappina, A. Becker, and A. Jaroń-Becker, *Phys. Rev. A* **76**, 063406 (2007).

[17] M. F. Ciappina, A. Becker, and A. Jaroń-Becker, *Phys. Rev. A* **78**, 063405 (2008).

[18] R. A. Ganeev, L. B. Elouga Bom, J. Abdul-Hadi, M. C. H. Wong, J. P. Brichta, V. R. Bhardwaj, and T. Ozaki, *Phys. Rev. Lett.* **102**, 013903 (2009).

[19] V. V. Afrosimov, A. A. Basalaev, M. N. Panov, and O. V. Smirnov, *Fullerences Nanotubes Carbon Nanostruct.* **12**, 485 (2004).

[20] S. C. O'Brien, J. R. Heat, R. F. Curl, and R. E. Smalley, *J. Chem. Phys.* **88**, 220 (1988).

[21] G. P. Zhang, X. Sun, and T. F. George, *Phys. Rev. B* **68**, 165410 (2003).

[22] V. R. Bhardwaj, P. B. Corkum, and D. M. Rayner, *Phys. Rev. Lett.* **91**, 203004 (2003).

[23] S. A. Trushin, W. Fuss, and W. E. Schmid, *J. Phys. B* **37**, 3987 (2004).

[24] I. V. Hertel, I. Shchatsinin, T. Laarmann, N. Zhavoronkov, H.-H. Ritze, and C. P. Schulz, *Phys. Rev. Lett.* **102**, 023003 (2009).

[25] I. Shchatsinin, H.-H. Ritze, C. P. Schulz, and I. V. Hertel, *Phys. Rev. A* **79**, 053414 (2009).

[26] L. Horváth and T. A. Beu, *Phys. Rev. B* **77**, 075102 (2008).

[27] T. A. Beu, L. Horváth, and I. Ghisoiu, *Phys. Rev. B* **79**, 054112 (2009).

[28] R. E. Allen, *Phys. Rev. B* **50**, 18629 (1994).

[29] M. Graf and P. Vogl, *Phys. Rev. B* **51**, 4940 (1995).

[30] J. S. Graves and R. E. Allen, *Phys. Rev. B* **58**, 13627 (1998).

[31] R. E. Allen, *Phys. Rev. B* **78**, 064305 (2008).

[32] M. Elstner, D. Porezag, G. Jungnickel, J. Elstner, M. Haugk, T. Frauenheim, S. Suhai, and G. Seifert, *Phys. Rev. B* **58**, 7260 (1998).

[33] T. Frauenheim, G. Seifert, M. Elstner, T. Niehaus, C. Köhler, M. Amkreutz, M. Sternberg, Z. Hajnal, A. Di Carlo, and S. Suhai, *J. Phys.: Condens. Matter* **14**, 3015 (2002).

[34] O. F. Sankey and D. J. Niklewski, *Phys. Rev. B* **40**, 3979 (1989).

[35] Q. Zhang and V. Buch, *J. Chem. Phys.* **92**, 5004 (1990).

[36] M. E. Riley, M. E. Coltrin, and D. J. Diestler, *J. Chem. Phys.* **88**, 5934 (1988).

[37] J. P. Perdew, K. Burke, and M. Ernzerhof, *Phys. Rev. Lett.* **77**, 3865 (1996).

[38] S. B. Fleischer, B. Pevzner, D. J. Dougherty, H. J. Zeiger, G. Dresselhaus, M. S. Dresselhaus, E. P. Ippen, and A. F. Hebard, *Appl. Phys. Lett.* **71**, 2734 (1997).

[39] B. Torralva, T. A. Niehaus, M. Elstner, S. Suhai, T. Frauenheim, and R. E. Allen, *Phys. Rev. B* **64**, 153105 (2001).

[40] V. R. Bhardwaj, P. B. Corkum, and D. M. Rayner, *Phys. Rev. Lett.* **91**, 203004 (2003).